\documentclass[12pt]{article}
\usepackage{amssymb,amsmath,amsthm}
\usepackage{hyperref}
\usepackage{array,longtable}

\usepackage[dvips]{graphicx}
\usepackage[dvips]{color}

\usepackage{wrapfig}
\usepackage{epsfig}
\makeatletter
   \newcommand\tabcaption{\def\@captype{table}\caption}
\makeatother

\usepackage[mathscr]{eucal}

\newcommand{\be}{\begin{equation}}
\newcommand{\ee}{\end{equation}}

\newcolumntype{V}{>{$}m{4cm}<{$}}
\newcolumntype{C}{>{$}c<{$}}
\newcolumntype{L}{>{$}l<{$}}
\newcolumntype{R}{>{$}r<{$}}

\allowdisplaybreaks[3]

\begin{document}
\thispagestyle{empty}

\vspace*{5.0ex}

\centerline{\huge \textsc{NS Matter Sliver}}


\vspace*{8.0ex}

\centerline{\large \rm I.Ya.~Aref'eva$^a$,
 A.A.~Giryavets$^b$ and P.B.~Medvedev$^c$}

\vspace*{7.0ex}

\centerline{\large \it ~$^a$Steklov Mathematical Institute,
 Russian Academy of Sciences,}
\centerline{\large \it Gubkin st. 8, Moscow, Russia, 117966 }

\centerline{\texttt{arefeva@mi.ras.ru}}

\vspace*{2.5ex}

\centerline{\large \it $^b$Faculty of Physics, Moscow State
University,} \centerline{\large \it Moscow, Russia, 119899 }

\centerline{ \texttt{alexgir@mail.ru}}

\vspace*{2.5ex}

\centerline{\large \it $^c$Institute of Theoretical and
Experimental Physics,} \centerline{\large \it B.Cheremushkinskaya
st. 25, Moscow, 117218}

\centerline{\texttt{medvedev@heron.itep.ru}}

\vspace*{10.0ex}

\centerline{\bf Abstract}
\medskip
Using algebraic  methods the Neveu-Schwarz fermionic matter sliver
is constructed. Inspirited by the wedge algebra, two equations for
the sliver, linear and quadratic, are considered. It is shown that
both equations give the same nontrivial answer. The sliver is
considered also using CFT methods where it is defined as the limit
of the wedge states in the NS sector of the superstring. \vfill \eject

\baselineskip=16pt

\newpage
\tableofcontents

\section{Introduction}
\setcounter{equation}{0}

The cubic open string field theory around the tachyon vacuum,
the vacuum string field theory (VSFT) proposed in \cite{F2}
 is investigated now very intensively
\cite{zwiebach}-\cite{0112169}. VSFT action has the same form as
the original Witten SFT action \cite{Witten}
\begin{equation}
S[\Phi]=\frac{1}{g_{o}^2}\left[\frac12 \int \Phi \star Q\Phi+
\frac13 \int \Phi \star \Phi \star\Phi\right], \label{WA}
\end{equation}
but with a new differential operator $\mathcal{Q}$ (for a review of
SFT see \cite{0102085,0109182,ABGKM}). VSFT is obtained by a shift
in the original Witten SFT action to the tachyon vacuum. The
absence of physical open string excitations around the tachyon
vacuum \cite{sen-con}, \cite{Hata1,Hata2,Taylor0103085,0105024}
motivates a suggestion \cite{F2} that after some field
redefinition $\mathcal{Q}$ can be written as a pure ghost
operator. If this conjecture is true it is worth to search for
solutions of VSFT equation of motion
\begin{equation}
\label{SEQ}
{\cal Q}\Phi +\Phi\star\Phi =0
\end{equation}
in the factorized form $\Phi =\Xi_{matter}\otimes \Phi_{ghost}$,
where $\Xi_{matter}$ satisfies a
projector-like equation:
\begin{equation}
\Xi_{matter}=\Xi_{matter}\star~\Xi_{matter}. \label{proj}
\end{equation}
An equation similar to (\ref{proj}) has appeared in a construction
of solitonic solutions in noncommutative field theories in the
large non-commutativity limit \cite{GMS}.

A way to solve projection equation (\ref{proj}) has been proposed
by Rastelli and Zwiebach \cite{zwiebach}. They have constructed a solution to
(\ref{proj}) as the $n\rightarrow\infty$ limit of
the wedge states $|n\rangle$. The wedge states are defined
on CFT language and they satisfy the algebra
\begin{equation}
|n\rangle\star |m\rangle=|n+m-1\rangle. \label{wedge-al}
\end{equation}
From algebra (\ref{wedge-al}) it immediately follows that $|\infty\rangle$,
the so-called sliver state, satisfies (\ref{proj}).

A solution of \eqref{proj} has been also constructed in
the oscillator formalism
by Kostelecky and Potting \cite{Kostelecky-Potting}. The
state $|\infty\rangle$ has been identified with the solution
\cite{Kostelecky-Potting} numerically \cite{RSZF} and later on by
direct calculations \cite{Japan2}. The sliver state and its
generalizations are studied in \cite{zwiebach}-\cite{0112169}.
 The sliver state is a special one in a sense that it can
be defined in an arbitrary boundary CFT \cite{zwiebach,rsz-4}. The
projector-like form of eq.(\ref{proj}) gives a new impulse for the
development of the half-string formalism \cite{GT01,GT02,rsz-3},
which drastically simplifies Witten's $\star$-product (see review
 \cite{ABGKM}). The previous consideration deals only with the
 bosonic SFT.

The subject of the present paper is a search for the sliver state
for the fermionic string. Open fermionic string in the NSR formalizm has a
tachyon in GSO$-$ sector that leads to a classical instability of
the perturbative vacuum in the theory without supersymmetry. It
has been proposed  \cite{sen-con} to interpret the tachyon
condensation in GSO$-$ sector of the NS string as a decay of unstable
non-BPS D-brane. By using the level truncation scheme it was shown
that the tachyon potential in the open cubic NS SFT has a
nontrivial minimum. In the framework of the Sen
interpretation the vacuum energy of the open NS GSO$-$ string
cancels the tension of the unstable space-filling 9-brane. This
cancellation has been checked \cite{ABKM}.

The cubic action for the NS GSO$+$ string field \cite{AMZ1,PTY}
can be be easily generalized to both GSO$+$ and GSO$-$ sectors
 \begin{equation}
S[\mathcal{A}]=\frac{1}{g_{o}^2}\left[\frac12 \int ^{\prime}
\mathcal{A} \star Q\mathcal{A}+ \frac13 \int
^{\prime}\mathcal{A} \star \mathcal{A} \star\mathcal{A}\right],
\label{action0}
\end{equation}
where $g_o$ is a coupling constant and the NS string
integral is modified accounting the "measure" $Y_{-2}$:
$\int^{\prime}=\int Y_{-2}$. Here  $Y_{-2}$ is a double-step
inverse picture changing operator. $\mathcal{A}$ describes both
GSO$+$ and GSO$-$ fields. As in the bosonic case one can
assume that VSSFT (vacuum super string filed theory) is described
by the cubic action (\ref{action0})
with a pure ghost $\mathcal{Q}$. Hence, there is
a reason to search for solutions of the VSSFT equation of motion
\begin{equation}
\label{SEQ2} {\cal Q}\mathcal{A} +\mathcal{A}\star\mathcal{A} =0
\end{equation}
in the factorized form
\begin{equation}
\label{fSEQ} \mathcal{A}
=\Xi^{X}\otimes\Xi^{\psi}\otimes\mathcal{A}_{ghost}.
\end{equation}

Note that the VSSFT equation of motion in the -1 picture \cite{Witten} would have the form
\begin{equation}
{\cal Q}\mathcal{A}+\mathrm{X}\mathcal{A}\star\mathcal{A}=0.
\label{W-eq}
\end{equation}
Because of the insertion of the picture-changing operator $X$
mixing ghost and matter
fields solutions of \eqref{W-eq} do not admit the factorized form.

The purpose of the present paper is a construction of the solution
of fermionic projector-like equation
\begin{equation}
|\Xi^{\psi}\rangle\star|\Xi^{\psi}\rangle=
|\Xi^{\psi}\rangle.\label{intr-sliv-eqXIXI-intr}
\end{equation}
Inspirited by the bosonic wedge algebra \eqref{wedge-al} we algebraically
construct a solution of the following equation
\begin{equation}
|\Xi^{\psi}\rangle\star|0\rangle=|\Xi^{\psi}\rangle,\label{intr-sliv-eqXI0}
\end{equation}
where $|0\rangle$ is a vacuum in the fermionic sector.
We show that \eqref{intr-sliv-eqXI0} and (\ref{intr-sliv-eqXIXI-intr})
have the same nontrivial solutions.

The NS matter sliver is constructed also using CFT methods.
 Numerical calculations give the remarkable  evidence that two
definitions of the sliver, conformal and algebraic, are
identical.

\section{Preliminary}
\label{sec:P}
\setcounter{equation}{0}

Fermionic coordinates in the Neveu-Schwarz sector are defined
for $\sigma\in[0,\pi]$ and are given by
\begin{equation}
\psi_{\pm}^{\mu}(\sigma)=\sum_{r=n+\frac{1}{2}}e^{\pm\imath
r\sigma}\psi_{r}^{\mu}.
\end{equation}
They satisfy the following anticommutation relations
\begin{equation}
\{\psi^{\mu}_{r},\psi^{\nu}_{s}\}=\delta_{r+s,\,0}g^{\mu\nu}.
\end{equation}
The fermionic squeezed states have the form
\begin{equation}
|S\rangle=\exp(\frac{1}{2}\psi^{\dag}_{r}S_{rs}\psi^{\dag}_{s})|0\rangle,
\label{sliv-sqv}
\end{equation}
One has
\begin{multline}
\langle
0|\exp(\frac{1}{2}\psi_{r}S_{sr}\psi_{s})\exp(\psi^{\dag}_{r}\mu_{r}+\frac{1}{2}\psi^{\dag}_{r}V_{rs}\psi^{\dag}_{s})|0\rangle\\
=\det(1+S_{rl}V_{ls})^{1/2}\exp(\frac{1}{2}\mu_{r}(1+S_{rl}V_{lk})^{-1}S_{ks}\mu_{s}),
\label{sqv}
\end{multline}
where $\mu_{r}$ are the anticommuting parameters.
The state $\langle A|$ is the BPZ conjugate of the state
$|A\rangle$ and is given by
\begin{equation}
~_{1}\langle A|= ~_{12}\langle R|A\rangle_{2},
\end{equation}
where $~_{12}\langle R|$ is the reflector \cite{GrJe3}
\begin{equation}
~_{12}\langle R|= ~_{12}\langle 0|\exp(-\imath
\psi^{1}_{r}(-)^{r}\psi^{2}_{r}). \label{ver-refl}
\end{equation}
The BPZ conjugate of \eqref{sliv-sqv} is
\begin{equation}
\langle S|=\langle
0|\exp(-\frac{1}{2}\psi_{r}C_{rl}S_{lk}C_{ks}\psi_{s}),
\end{equation}
where $C_{rs}=(-1)^{r}\delta_{rs}$ and $C^2=-1$.

Following Gross and Jevicki \cite{GrJe,GrJe3} we represent the
star product of the states $|A\rangle$ and $|B\rangle$ in the
Neveu-Schwarz sector of the matter fermionic SFT as
\begin{equation}
|A\rangle\star|B\rangle_{1}= ~_{2}\langle A|~_{3}\langle
B|V_{3}\rangle,
\end{equation}
where the three string vertex is given by
\begin{equation}
|V_{3}\rangle=\exp(\frac{1}{2}\psi^{a\,\dag}_{r}V^{ab}_{rs}\psi^{b\,\dag}_{s})|0\rangle_{123},
\label{ver-ver}
\end{equation}
where
\begin{subequations}
\label{ver-IUCUC}
\begin{gather}
V^{a\,a}=\frac{1}{3}(I+U+CUC),\\
V^{a\,a+1}=\frac{1}{3}(I+\alpha U+\alpha^{*} CUC),\\
V^{a\,a-1}=\frac{1}{3}(I+\alpha^{*} U+\alpha CUC),
\end{gather}
\end{subequations}
and $\alpha=e^{2\pi\imath/3}$. The three string
vertex has the cyclic
property $V^{ab}=V^{(a+1)(b+1)}$, where indexes $a,b$ are defined
mod(3). Matrices $I$ and $U$ are calculated in appendix A and are
given by
\begin{equation}
I=\frac{\tilde{F}}{1-F},\qquad
U=-\frac{\tilde{F}-\sqrt{3}C}{2+F},\qquad
CUC=-\frac{\tilde{F}+\sqrt{3}C}{2+F}.
\end{equation}
Here $F$ and $\tilde{F}$ are hermitian matrices \cite{GrJe3}
\begin{gather}
F_{rs}=-\frac{2}{\pi}\frac{\imath^{r-s}}{r+s},\quad
r=s\text{ mod}(2),\\
\tilde{F}_{rs}=\frac{2}{\pi}\frac{\imath^{r+s}}{s-r},\quad
r=s+1\text{ mod}(2)
\end{gather}
with the following properties
\begin{gather}
F^{2}-\tilde{F}^{2}=1,\quad [F,\tilde{F}]=0,\\
CFC=-F,\quad F^{T}=F,\quad C\tilde{F}C=\tilde{F},\quad
\tilde{F}^{T}=-\tilde{F}.
\end{gather}
In terms of $F$, $\tilde{F}$ and $C$ the three string vertex is
given by
\begin{subequations}
\begin{gather}
V^{11}=\frac{F\tilde{F}}{(1-F)(2+F)},\\
V^{12}=\frac{\tilde{F}+\imath C(1-F)}{(1-F)(2+F)},\\
V^{21}=\frac{\tilde{F}-\imath C(1-F)}{(1-F)(2+F)}.
\end{gather}
\end{subequations}
We will use the following notation
\begin{equation}
M_{ab}=CV^{ab}.
\end{equation}
One has the following properties for $M_{ab}$
\begin{subequations}
\label{matrix-m-prop}
\begin{gather}
M_{12}+M_{21}+M_{11}=CI,\\
[M_{ab},M_{cd}]=0,\quad \forall \;\; a, b, c, d,\\
M_{12}M_{21}=M_{11}^{2}-CI^{-1}M_{11}.
\end{gather}
\end{subequations}

\section{Algebraic Construction}
\label{sec:AC}
\setcounter{equation}{0}

\subsection{Linear equation}
Let us assume that $|\Xi^{\psi}\rangle$ has a squeezed form \eqref{sliv-sqv}
\begin{equation}
|\Xi^{\psi}\rangle=\mathcal{N}^{10}\exp(\frac{1}{2}\psi^{\dag}_{r}S_{rs}\psi^{\dag}_{s})|0\rangle,\label{sliv-sqv-al}
\end{equation}
and let us solve the following equation
\begin{equation}
|\Xi^{\psi}\rangle\star|0\rangle=|\Xi^{\psi}\rangle.\label{sliv-eqXI0}
\end{equation}
Here $\mathcal{N}$ is the normalization factor to be
specified later. Using \eqref{sqv} one gets
the following equation
\begin{equation}
M_{21}(1-TM_{11})^{-1}TM_{12}+M_{11}=T. \label{sliv-eqS0}
\end{equation}
Here we denote $T=CS$.  Assuming the
following commutation relations
\begin{equation}
[T,M_{ab}]=0,\quad \forall\;\; a,b,
\end{equation}
and using the properties \eqref{matrix-m-prop}, equation
\eqref{sliv-eqS0} can be rewritten as
\begin{equation}
T^{2}M_{11}-T(1+CI^{-1}M_{11})+M_{11}=0.\label{sliv-T-eq}
\end{equation}
An explicit solution of this equation is
\begin{equation}
T=\frac{1+CI^{-1}M_{11}-\sqrt{(1+CI^{-1}M_{11})^{2}-4M_{11}^{2}}}{2M_{11}}.
\label{sliv-m}
\end{equation}
We choose the minus sign for the square root in \eqref{sliv-m}. We
will check later this choice of the sign.

\subsection{Quadratic equation}
Let us now consider the projector-like equation
\begin{equation}
|\Xi^{\psi}\rangle\star|\Xi^{\psi}\rangle=|\Xi^{\psi}\rangle,\label{sliv-eqXIXI}
\end{equation}
where $|\Xi^{\psi}\rangle$ has the form \eqref{sliv-sqv-al}.
Using \eqref{sqv} one
gets from \eqref{sliv-eqXIXI} the following equation
\begin{equation}
(M_{12},M_{21})\left(1-T\begin{pmatrix}
  M_{11} & M_{12} \\
  M_{21} & M_{22}
\end{pmatrix} \right)^{-1}\begin{pmatrix}
  TM_{21} \\
  TM_{12}
\end{pmatrix}+M_{11}=T.
\end{equation}
This equation can be rewritten in the form
\begin{equation}
(T-CI)(T^{2}M_{11}-T(1+CI^{-1}M_{11})+M_{11})=0.\label{sliv-eqSS}
\end{equation}
We get a trivial solution $T=CI$. It corresponds to the identity state
$|I\rangle$. The identity state is the identity of the star algebra
and has the form \cite{GrJe3}
\begin{equation}
|I\rangle=\exp(\frac{1}{2}\psi^{\dag}_{r}I_{rs}\psi^{\dag}_{s})|0\rangle.
\end{equation}
Up to the factor $(T-CI)$, equation \eqref{sliv-eqSS} is the same as
equation \eqref{sliv-eqS0}.

Finally, one gets the following expression for the NS matter sliver
\begin{equation}
|\Xi^{\psi}\rangle=\mathcal{N}^{10}\exp(\frac{1}{2}\psi^{\dag}_{r}S_{rs}\psi^{\dag}_{s})|0\rangle,
\end{equation}
where one can find normalization factor from eq.
\eqref{sliv-eqXIXI}
\begin{multline}
\mathcal{N}=\det((1-TM_{11})^{2}-T^{2}M_{12}M_{21})^{-1/2}\\
=\det(1-CI^{-1}M_{11}+T(CI^{-1}-2M_{11}+I^{-2}M_{11}))^{-1/2}.
\end{multline}
Using the normalization $\langle 0|0\rangle=1$,
one can now write the norm of the NS sliver state
\begin{equation}
\langle\Xi^{\psi}|\Xi^{\psi}\rangle=\mathcal{N}^{20}\det(1-S^{2})^{5}.\label{sliv-norm}
\end{equation}

\section{NS Wedge States}
\label{sec:NSWS}
\setcounter{equation}{0}

A generalization of the bosonic wedge states
\cite{zwiebach,RSZF,rsz-4} to the fermionic wedge states is
straightforward.
Wedge states $|n\rangle$ are defined by
\begin{equation}
\langle n|\phi^{\psi}\rangle=\langle f_{n}\circ\phi^{\psi}(0)\rangle,
\end{equation}
where $|\phi^{\psi}\rangle$ is an arbitrary state which belongs to
the fermionic subspace, $f_{n}\circ\phi^{\psi}(\xi)$ denotes the
conformal transform of $\phi^{\psi}(\xi)$ and $f_{n}(\xi)$ is
the same as in the bosonic case, i.e.
\begin{gather}
f_{n}(\xi)=\frac{n}{2}\tan\left(\frac{2}{n}\tan^{-1}\xi\right).\label{wedge-map}
\end{gather}
The wedge
state multiplication rule can be shown to be \cite{zwiebach}
\begin{equation}
|n\rangle\star |m\rangle=|n+m-1\rangle.
\end{equation}
The wedge state with $n=1$ corresponds to the identity of the star
algebra and with $n=2$ corresponds to the vacuum.

 Taking the limit
$n\rightarrow\infty$ in \eqref{wedge-map} one derives the
conformal map for the sliver state $|\infty\rangle$
\begin{equation}
w(\xi)=\tan^{-1}(\xi).\label{sliv-map}
\end{equation}
Following \cite{GrJe3,l'Clare,l'Clare2} one gets the explicit
formula for the state $|\Lambda\rangle$ corresponding to a
conformal map $\lambda(\xi)$
\begin{equation}
|\Lambda\rangle\propto\exp(\frac{1}{2}\psi^{\dag}_{r}\Lambda_{rs}\psi^{\dag}_{s})|0\rangle,
\end{equation}
\begin{equation}
\Lambda_{rs}=\oint\frac{d\xi}{2\pi\imath}\oint\frac{d\xi'}{2\pi\imath}\xi^{-r-\frac{1}{2}}\xi'{}^{-s-\frac{1}{2}}
\left(\frac{\partial
\lambda(\xi)}{\partial\xi}\right)^{\frac{1}{2}}\frac{1}{\lambda(\xi)-\lambda(\xi')}\left(\frac{\partial
\lambda(\xi')}{\partial\xi'}\right)^{\frac{1}{2}}.\label{state-map}
\end{equation}
Here $\oint$ denotes the contour integration around the origin.
Using the sliver conformal map \eqref{sliv-map} one gets
\begin{equation}
\left(\frac{\partial
w(\xi)}{\partial\xi}\right)^{\frac{1}{2}}\frac{1}{w(\xi)-w(\xi')}\left(\frac{\partial
w(\xi')}{\partial\xi'}\right)^{\frac{1}{2}}=\frac{2\imath}{\sqrt{1+\xi^2}\,\sqrt{1+\xi'{}^2}}\ln\left(\frac{(1+i\xi)(1-i\xi')}{(1-i\xi)(1+i\xi')}\right).
\end{equation}
So the conformal sliver
$|\tilde{\Xi}^{\psi}\rangle\equiv|\infty\rangle$ is defined as
\begin{equation}
|\tilde{\Xi}^{\psi}\rangle=\tilde{\mathcal{N}}^{10}\exp(\frac{1}{2}\psi^{\dag}_{r}\tilde{S}_{rs}\psi^{\dag}_{s})|0\rangle,
\end{equation}
\begin{equation}
\tilde{S}_{rs}=\oint\frac{d\xi}{2\pi\imath}\oint\frac{d\xi'}{2\pi\imath}\xi^{-r-\frac{1}{2}}\xi'{}^{-s-\frac{1}{2}}
\frac{2\imath}{\sqrt{1+\xi^2}\,\sqrt{1+\xi'{}^2}}\ln\left(\frac{(1+i\xi)(1-i\xi')}{(1-i\xi)(1+i\xi')}\right).
\end{equation}
The matrix $\tilde{S}_{rs}$ can be calculated explicitly. Only
coefficients with $r+s=even$ differ from zero.

\newpage
\section{Comparison}
\label{sec:C}
\setcounter{equation}{0}

Results of calculation for the algebraic sliver $|\Xi^{\psi}\rangle$
that are presented on Table \ref{table} are in accord with the
results of calculations for the conformal sliver
$|\tilde{\Xi}^{\psi}\rangle$ presented below
\begin{subequations}
\begin{align}
\tilde{S}_{\frac{1}{2}\frac{3}{2}}&=\frac{1}{6}\approx
0.16667,\qquad
&\tilde{S}_{\frac{1}{2}\frac{7}{2}}&=-\frac{43}{360}\approx
-0.11944,\\
\tilde{S}_{\frac{1}{2}\frac{11}{2}}&=\frac{1459}{15120}\approx
0.09649,\qquad
&\tilde{S}_{\frac{5}{2}\frac{3}{2}}&=-\frac{1}{40}=-0.02500,\\
\tilde{S}_{\frac{5}{2}\frac{7}{2}}&=\frac{239}{7560}\approx
0.03161,\qquad
&\tilde{S}_{\frac{5}{2}\frac{11}{2}}&=-\frac{18947}{604800}\approx
-0.03133.
\end{align}
\label{num-conf}
\end{subequations}
We see a conspicuous agreement between $S_{rs}$ and $\tilde{S}_{rs}$.
This gives a convincing evidence that two descriptions of
the NS sliver, algebraic and conformal, are identical.
Normalization factors $\mathcal{N}$
and $\tilde{\mathcal{N}}$ should be equal since these two states
are normalized identically.
\begin{table}[h]
\begin{center}\def\st{\vrule height 3ex width 0ex}
\begin{tabular}{|l|l|l|l|l|l|l|} \hline
$L$ & $S_{\frac{1}{2}\frac{3}{2}}$ & $S_{\frac{1}{2}\frac{7}{2}}$
& $S_{\frac{1}{2}\frac{11}{2}}$ & $S_{\frac{5}{2}\frac{3}{2}}$ &
$S_{\frac{5}{2}\frac{7}{2}}$ & $S_{\frac{5}{2}\frac{11}{2}}$
\st\\[1ex]
\hline \hline 30  & 0.13744 &  $-$0.09233 & 0.07098 & $-$0.01261 &
0.01996 & $-$0.02026
\st\\[1ex]
\hline 60 & 0.14667 & $-$0.09950 & 0.07696 & $-$0.01658 & 0.02305
& $-$0.02286
\st\\[1ex]
\hline 100 & 0.15306 & $-$0.10471 & 0.08146 & $-$0.01940 & 0.02535
& $-$0.02484
\st\\[1ex]
\hline 130 & 0.15616 &  $-$0.10729 & 0.08372 & $-$0.02078 &
0.02650 & $-$0.02585
\st\\[1ex]
\hline Exact & 0.16667 &  $-$0.11944 & 0.09649 & $-$0.02500 &
0.03161 & $-$0.03133
\st\\[1ex]

\hline
\end{tabular}
\end{center}
\caption{Numerical results for the elements of the matrix
$S_{rs}$. We compute $S_{rs}$ by restricting the indices $r,s$ of
$V^{11}_{rs}$ and $I^{-1}_{rs}$ to be less or equal to $L$ so that $V^{11}$
and $I^{-1}$ are $L\times L$ matrices, and then using equation
\eqref{sliv-m}. In the last line the results \eqref{num-conf} are presented.}
\label{table}
\end{table}


\section{Conclusion}
\setcounter{equation}{0}

In this paper we have constructed the fermionic matter part of the
VSSFT equations. The VSSFT action (\ref{action0}) on
a factorized solution of the equations of motion is:
\begin{equation}
\label{val-ac}
S|_{\Xi}=-\frac{1}{6g_{o}^2}\langle \Xi^{X}|\Xi^{X}\rangle
\langle\Xi^{\psi}|\Xi^{\psi}\rangle \langle\!\langle
Y_{-2}|\mathcal{A}_{gh},\mathcal{Q}\mathcal{A}_{gh}\rangle\!\rangle.
\end{equation}

Eq. (\ref{val-ac}) shows that the matter fermionic part
drop out from the ratio of tensions of D-$(9-k)$
and D-$(8-k)$ branes. Indeed,
the ratio of the actions associated with the solutions representing
non-BPS
D-branes of different dimensions is given by
\begin{equation}
\frac{S|_{\Xi'}}{S|_{\Xi}}=\frac{\langle
\Xi'{}^{X}|\Xi'{}^{X}\rangle}{\langle
\Xi^{X}|\Xi^{X}\rangle},\label{ratio}
\end{equation}
since these D-branes have the same matter fermionic and ghost
parts. The ghost and matter fermionic parts drop out both for momentum
dependent and momentum independent solutions. In \cite{RSZF} it is
argued that in the bosonic case the ratio of tensions of D-$(25-k)$
and D-$(24-k)$ branes  given by  \eqref{ratio} is equal to
$1/2\pi$. A similar result one gets for the ratio of tensions in
the VSSFT
\begin{equation}
\frac{\tau_{9-k}}{\tau_{8-k}}=\frac{1}{2\pi}.
\label{rel}
\end{equation}
Assuming that  the tachyonic kinks describe lower
dimensional non-PBPS D-branes we do expect the relation (\ref{rel}).

{\bf Added note:}
After this paper was submitted to hep-th, the  paper \cite{Marino} appeared.
In this paper the fermion sliver is independently constructed.
\section*{Acknowledgments.}

We would like to thank
D.M.Belov  and A.S.Koshelev for discussions.
This work was supported in part
by RFBR grant 99-01-00166, RFBR grant for leading scientific
schools and  by INTAS grant
99-0590.

\appendix

\section{NS String Overlaps}
The three string overlaps \cite{GrJe3} are
\begin{gather}
\psi^{i}_{+}(\sigma )=\imath \psi^{i-1}_{+}(\pi-\sigma),\quad 0\leq\sigma\leq\frac{\pi}{2},\nonumber\\
\psi^{i}_{-}(\sigma )=-\imath \psi^{i-1}_{-}(\pi-\sigma),\quad
0\leq\sigma\leq\frac{\pi}{2}
 \label{x1}
\end{gather}
and
\begin{gather}
\psi^{i-1}_{+}(\sigma )=-\imath \psi^{i}_{+}(\pi-\sigma),\quad \frac{\pi}{2}\leq\sigma\leq\pi,\nonumber\\
\psi^{i-1}_{-}(\sigma )=\imath \psi^{i}_{-}(\pi-\sigma),\quad
\frac{\pi}{2}\leq\sigma\leq\pi . \label{x2}
\end{gather}
In terms of $\psi^{i}(\sigma)$ overlaps can be rewritten in the
following way
\begin{equation}
-\pi\leq\sigma\leq \pi, ~~~~ \psi ^{i}(\sigma )=
\begin{cases}
i\psi^{i+1}(-\sigma -\pi),\quad-\pi\leq\sigma\leq -\frac{\pi}{2},\\
-i\psi^{i-1}(-\sigma -\pi),\quad -\frac{\pi}{2}\leq\sigma\leq 0,\\
i\psi^{i-1}(\pi -\sigma ),\quad 0\leq\sigma\leq\frac{\pi}{2},\\
-i\psi^{i+1}(\pi -\sigma ),\quad\frac{\pi}{2}\leq\sigma\leq\pi.
\end{cases}
\end{equation}
Introducing $Z_{3}$ Fourier coordinates
\begin{subequations}
\begin{align}
\Psi^{1}&=\frac{1}{\sqrt{3}}(\psi^{1}_{+}+\psi^{2}_{+}+\psi^{3}_{+}),\\
\Psi^{2}&=\frac{1}{\sqrt{3}}(\psi^{1}_{+}+\alpha\psi^{2}_{+}+\alpha^{*}\psi^{3}_{+})\equiv\Psi,\\
\Psi^{3}&=\frac{1}{\sqrt{3}}(\psi^{1}_{+}+\alpha^{*}\psi^{2}_{+}+\alpha\psi^{3}_{+})\equiv\Bar{\Psi},
\end{align}
\end{subequations}
one gets the identity overlap for $\Psi^{1}(\sigma)$ and
the following overlaps for $\Psi(\sigma)$ and $\Bar{\Psi}(\sigma)$  \cite{GrJe3}
\begin{equation}
-\pi\leq\sigma\leq \pi,~~~~ \Psi(\sigma )=
\begin{cases}
i\alpha^{\ast}\Psi(-\sigma -\pi),\quad-\pi\leq\sigma\leq -\frac{\pi}{2},\\
-i\alpha\Psi(-\sigma -\pi),\quad -\frac{\pi}{2}\leq\sigma\leq 0,\\
i\alpha\Psi(\pi -\sigma ),\quad 0\leq\sigma\leq\frac{\pi}{2},\\
-i\alpha^{\ast}\Psi(\pi -\sigma
),\quad\frac{\pi}{2}\leq\sigma\leq\pi,
\end{cases}
\end{equation}
\begin{equation}
-\pi\leq\sigma\leq \pi,~~~~ \Bar{\Psi}(\sigma )=
\begin{cases}
i\alpha\Bar{\Psi}(-\sigma -\pi),\quad-\pi\leq\sigma\leq -\frac{\pi}{2},\\
-i\alpha^{\ast}\Bar{\Psi}(-\sigma -\pi),\quad -\frac{\pi}{2}\leq\sigma\leq 0,\\
i\alpha^{\ast}\Bar{\Psi}(\pi -\sigma ),\quad 0\leq\sigma\leq\frac{\pi}{2},\\
-i\alpha\Bar{\Psi}(\pi -\sigma
),\quad\frac{\pi}{2}\leq\sigma\leq\pi.
\end{cases}
\end{equation}
Overlap equations for $\Psi(\sigma)$ can be rewritten in the
component form
\begin{gather}
\Psi_{r}=-\frac{1}{2}F_{rs}\Psi_{s}-\frac{1}{2}(\tilde{F}_{rs}-\sqrt{3}C)\Psi_{-s},\\
\Psi_{-r}=\frac{1}{2}(\tilde{F}_{rs}-\sqrt{3}C)\Psi_{s}+\frac{1}{2}F_{rs}\Psi_{-s}.
\end{gather}
We search the three string vertex in the following form
\begin{equation}
|V_{3}\rangle=\exp\left(\frac{1}{2}\Psi^{1\,\dag}I\Psi^{1\,\dag}
+\Bar{\Psi}^{\dag}U\Psi^{\dag}\right)|0\rangle_{123}.
\end{equation}
One finds the following equations for $U$
\begin{gather}
U=-\frac{1}{2}FU-\frac{1}{2}(\tilde{F}-\sqrt{3}C),\\
1=\frac{1}{2}(\tilde{F}-\sqrt{3}C)U+\frac{1}{2}F.
\end{gather}
They give solutions
\begin{gather}
U=-\frac{\tilde{F}-\sqrt{3}C}{2+F},\\
U=\frac{2-F}{\tilde{F}-\sqrt{3}C}.
\end{gather}
These two solutions are equivalent. Solving overlaps for
$\Bar{\Psi}(\sigma)$ one gets the following equivalent solutions
for $CUC$
\begin{gather}
CUC=-\frac{\tilde{F}+\sqrt{3}C}{2+F},\\
CUC=\frac{2-F}{\tilde{F}+\sqrt{3}C}.
\end{gather}



\section{Conformal Definition of Vertex}
Here we present some conformal formulae of \cite{GrJe3}. Formula
for the state corresponding to the conformal map $\lambda(\xi)$
\eqref{state-map} can be rewritten in terms of $x=i\xi$
\begin{equation}
\Lambda_{rs}=\imath^{r+s}\oint\frac{dx}{2\pi\imath}\oint\frac{dx'}{2\pi\imath}x^{-r-\frac{1}{2}}x'{}^{-s-\frac{1}{2}}
\left(\frac{\partial \lambda(x)}{\partial
x}\right)^{\frac{1}{2}}\frac{1}{\lambda(x)-\lambda(x')}\left(\frac{\partial
\lambda(x')}{\partial x'}\right)^{\frac{1}{2}}.
\end{equation}
For the identity state conformal map has the following form
\begin{equation}
\lambda(x)=\left(i\frac{1+x}{1-x}\right)^{2}\equiv z^{2}.
\end{equation}
Using the generation function defined as
\begin{equation}
\left(\frac{1+x}{1-x}\right)^{\frac{1}{2}}=\sum_{n=0}^{\infty}u_{n}x^{n},\quad
u_{0}=u_{1}=1,\quad
u_{2n}=u_{2n+1}=\frac{\left(\frac{2n-1}{2}\right)!}{n!},\; n\geq1.
\end{equation}
one can write the identity in the simple form
\begin{equation}
I_{rs}=\imath^{r+s}\left(\frac{I^{+}_{nm}}{r+s}+\frac{I^{-}_{nm}}{r-s}\right)=\imath^{r+s}
\begin{cases}
\left(\frac{-m}{n+m+1}-\frac{m}{n-m}\right)u_{n}u_{m},\quad n=\text{even},\; m=\text{odd},\\
\left(\frac{n}{n+m+1}-\frac{n}{n-m}\right)u_{n}u_{m},\quad
n=\text{odd},\; m=\text{even},
\end{cases}
\end{equation}
where $r=n+\frac{1}{2}$ and $s=m+\frac{1}{2}$.\\
Fermionic ghosts have the identity equal to $I^{-1}$
\begin{equation}
I^{-1}_{rs}=\imath^{r+s}\left(\frac{I^{+}_{nm}}{r+s}-\frac{I^{-}_{nm}}{r-s}\right)=\imath^{r+s}
\begin{cases}
\left(\frac{-m}{n+m+1}+\frac{m}{n-m}\right)u_{n}u_{m},\quad n=\text{even},\; m=\text{odd},\\
\left(\frac{n}{n+m+1}+\frac{n}{n-m}\right)u_{n}u_{m},\quad
n=\text{odd},\; m=\text{even}.
\end{cases}
\end{equation}
To calculate the fermionic vertex one has the following conformal
maps
\begin{equation}
w_{a}=w^{0}_{a}\left(\frac{1+x}{1-x}\right)^{\frac{2}{3}},\quad
w_{1}^{0}=e^{\imath\frac{\pi}{3}},\;
w_{2}^{0}=e^{-\imath\frac{\pi}{3}},\;w_{3}^{0}=e^{-\imath\pi}.
\end{equation}
Using these conformal maps and the generation function defined as
\begin{equation}
\left(\frac{1+x}{1-x}\right)^{\frac{1}{6}}=\sum_{n=0}^{\infty}g_{n}x^{n},\quad
g_{0}=1,\; g_{1}=\frac{1}{3},\;
g_{n+1}=\frac{1}{3(n+1)}g_{n}+\frac{n-1}{n+1}g_{n-1},\; n\geq1
\end{equation}
one gets the following formulae for the three string vertex
\begin{gather}
V^{a\,a}_{rs}=\frac{1}{3}I_{rs}+\imath^{r+s}\left[\frac{M^{+}_{nm}}{r+s}+\frac{M^{-}_{nm}}{r-s}\right],\\
V^{a\,a+1}_{rs}=\frac{1}{2}I_{rs}-\frac{1}{2}V^{aa}_{rs}+\frac{\sqrt{3}}{2}\imath^{r+s+1}\left[\frac{\Bar{M}^{+}_{nm}}{r+s}+\frac{\Bar{M}^{-}_{nm}}{r-s}\right],\\
V^{a\,a-1}_{rs}=\frac{1}{2}I_{rs}-\frac{1}{2}V^{aa}_{rs}-\frac{\sqrt{3}}{2}\imath^{r+s+1}\left[\frac{\Bar{M}^{+}_{nm}}{r+s}+\frac{\Bar{M}^{-}_{nm}}{r-s}\right],
\end{gather}
where the matrices $M^{\pm}$ and $\Bar{M}^{\pm}$ are defined as
follows
\begin{gather}
M^{+}_{nm}=-[(n+1)g_{n+1}(m+1)g_{m+1}-ng_{n}mg_{m}][(-)^{n}-(-)^{m}],\\
M^{-}_{nm}=-[ng_{n}(m+1)g_{m+1}-(n+1)g_{n+1}mg_{m}][(-)^{n}-(-)^{m}],\\
\Bar{M}^{+}_{nm}=[(n+1)g_{n+1}(m+1)g_{m+1}-ng_{n}mg_{m}][(-)^{n}+(-)^{m}],\\
\Bar{M}^{-}_{nm}=[ng_{n}(m+1)g_{m+1}-(n+1)g_{n+1}mg_{m}][(-)^{n}+(-)^{m}],\\
\frac{M_{rs}^{\pm}}{r\pm s}|_{r=s}=0,\qquad
\frac{\Bar{M}_{rs}^{+}}{r+s}|_{r=s}=\frac{(n+1)^{2}g_{n+1}^{2}-n^{2}g_{n}^{2}}{n+\frac{1}{2}}(-)^{n},\\
\frac{\Bar{M}_{rs}^{-}}{r-s}|_{r=s}=\frac{2}{3}\sum_{k=0}^{n}(-)^{n-k}g_{n-k}^{2}.
\end{gather}
This three string vertex can be rewritten in the form \eqref{ver-IUCUC}.

{\small


\begin{thebibliography}{99}

\bibitem{F2} L.~Rastelli, A.~Sen, B.~Zwiebach,
\textit{String field theory  around the tachyon vacuum},
hep-th/0012251.



\bibitem{zwiebach} L.~Rastelli and B.~Zwiebach,
\textit{Tachyon potentials, star products and universality}, JHEP
0109 (2001) 038, hep-th/0006240.


\bibitem{Kostelecky-Potting}
V.A.~Kostelecky and R.~Potting, \textit{Analytical construction of
a nonperturbative vacuum for the open bosonic string},
Phys.Rev.~D63 (2001) 046007, hep-th/0008252.


\bibitem{RSZF} L.~Rastelli, A.~Sen, B.~Zwiebach,
\textit{Classical Solutions in String Field Theory Around the
Tachyon Vacuum}, hep-th/0102112.


\bibitem{GT01} D.J.~Gross and W.~Taylor,
\textit{Split string field theory. I}, JHEP 0108, 009 (2001),
hep-th/0105059.


\bibitem{GT02} D.J.~Gross and W.~Taylor,
\textit{Split string field theory. II}, JHEP 0108, 010 (2001),
hep-th/0106036.


\bibitem{rsz-3} L.~Rastelli, A.~Sen and B.~Zwiebach,
\textit{Half-strings, projectors, and multiple D-branes in vacuum
string field theory}, JHEP 0111 (2001) 035, hep-th/0105058.


\bibitem{rsz-4} L.~Rastelli, A.~Sen and B.~Zwiebach,
\textit{Boundary CFT construction of D-branes in vacuum string
field theory}, JHEP 0111 (2001) 045, hep-th/0105168.


\bibitem{Okuyama}
T.~Kawano and K.~Okuyama, \textit{Open string fields as matrices},
JHEP 0106 (2001) 061, hep-th/0105129.


\bibitem{rsz-5} L.~Rastelli, A.~Sen and B.~Zwiebach,
\textit{Vacuum string field theory}, hep-th/0106010.


\bibitem{0105184}
J.R.~David, \textit{Excitations on wedge states and on the
sliver}, JHEP 0107 (2001) 024, hep-th/0105184.


\bibitem{0106242} D.J.~Gross, V.~Periwal,
\textit{String field theory, non-commutative Chern-Simons theory
and Lie algebra cohomology}, JHEP 0108 (2001) 008, hep-th/0106242.


\bibitem{Japan2}
K.~Furuuchi and K.~Okuyama, \textit{Comma vertex and string field
algebra}, JHEP 0109 (2001) 035, hep-th/0107101.


\bibitem{Matsuo} Y.~Matsuo,
\textit{BCFT and sliver state}, Phys.Lett. B513 (2001) 195-199,
hep-th/0105175.
\\
 Y.~Matsuo, \textit{Identity projector and D-brane in
string field theory}, Phys.Lett. B514 (2001) 407-412,
hep-th/0106027.
\\
 Y.~Matsuo, \textit{Projection operators and D-branes
in purely cubic open string field theory}, Mod.Phys.Lett. A16
(2001) 1811-1822, hep-th/0107007.


\bibitem{0108150} H.~Hata and T.~Kawano,
\textit{Open string states around a classical solution in vacuum
string field theory}, JHEP 0111 (2001) 038, hep-th/0108150.


\bibitem{0110124} I.~Kishimoto,
\textit{Some properties of string field algebra}, JHEP 0112 (2001)
007, hep-th/0110124.


\bibitem{0110136} P.~Mukhopadhyay,
\textit{Oscillator representation of the BCFT construction of
D-branes in vacuum string filed theory}, hep-th/0110136.


\bibitem{0110204} N.~Moeller,
\textit{Some exact results on the matter star-product in the
half-string formalism}, hep-th/0110204.


\bibitem{HataSliver2} H.~Hata and S.~Moriyama,
\textit{Observables as twist anomaly in vacuum string field
theory}, hep-th/0111034.


\bibitem{MooreTaylor} G.~Moore and W.~Taylor,
\textit{The singular geometry of the sliver}, hep-th/0111069.


\bibitem{0111087}  K.~Okuyama,
\textit{Siegel Gauge in Vacuum String Field Theory},
hep-th/0111087.


\bibitem{0111092} A.~Hashimoto and N.~Itzhaki,
\textit{Observables of String Field Theory}, hep-th/0111092.


\bibitem{rsz-6} D.~Gaiotto, L.~Rastelli, A.~Sen and B.~Zwiebach,
\textit{Ghost Structure and Closed Strings in Vacuum String Field
Theory}, hep-th/0111129.


\bibitem{rsz-7} L.~Rastelli, A.~Sen and B.~Zwiebach,
\textit{A Note on a Proposal for the Tachyon State in Vacuum
String Field Theory}, hep-th/0111153.


\bibitem{0111281} L.~Rastelli, A.~Sen and B.~Zwiebach,
\textit{Star Algebra Spectroscopy}, hep-th/0111281.


\bibitem{0112124} T.~Takahashi, S.~Tanimoto,
\textit{Dilaton Condensation in Cubic Open String Field Theory},
hep-th/0112124.


\bibitem{0112169} I.~Kishimoto, K.~Ohmori,
\textit{CFT Description of Identity String Field: Toward
Derivation of the VSFT Action}, hep-th/0112169.



\bibitem{Witten} E.~Witten,
\textit{Noncommutative geometry and string field theory},
Nucl.Phys. B268 (1986) 253;
\\
E.~Witten, \textit{Interacting field theory of open superstrings},
Nucl.Phys.~B276 (1986) 291.


\bibitem{0102085}
K.~Ohmori, \textit{A Review on Tachyon Condensation in Open String
Field Theories}, hep-th/0102085.


\bibitem{0109182} P. De Smet,
\textit{Tachyon Condensation: Calculations in String Field Theory
}, hep-th/0109182.


\bibitem{ABGKM} I.Ya.~Aref'eva, D.M.~Belov, A.A.~Giryavets, A.S.~Koshelev,
P.B.~Medvedev, \textit{Noncommutative Field Theories and
(Super)String Field Theories}, hep-th/0111208.



\bibitem{sen-con} A.~Sen,
\textit{Stable non BPS bound states of BPS D-branes}, JHEP 08 010 (1998),
hep-th/9805019.
\\
A.~Sen, \textit{Descent relations among bosonic
D-branes}, Int. J. Mod. Phys. A 14 (1999) 4061, hep-th/9902105.
\\
A.~Sen,
\textit{SO(32) spinors of type I and other solitons on brane-anti-brane pair},
JHEP 09 023 (1998), hep-th/9808141.
\\
A.~Sen, \textit{Non-BPS States and Branes in String Theory},
hep-th/9904207.






\bibitem{Hata1} H.~Hata, S.~Shinohara,
\textit{BRST Invariance of the Non-Perturbative Vacuum in Bosonic
Open String Field Theory},  JHEP 0009 (2000) 035, hep-th/0009105.


\bibitem{Hata2} H.~Hata, S.~Teraguchi,
\textit{Test of the Absence of Kinetic Terms around the Tachyon
Vacuum in Cubic String Field Theory}, JHEP 0105 (2001) 045,
hep-th/0101162.


\bibitem{Taylor0103085} I.~Ellwood, W.~Taylor,
\textit{Open string field theory without open strings}, Phys.Lett.
B512 (2001) 181, hep-th/0103085.


\bibitem{0105024} I.~Ellwood, B.~Feng, Y.-H.~He and N.~Moeller,
\textit{The Identity String Field and the Tachyon Vacuum}, JHEP
0107 (2001) 016, hep-th/0105024.



\bibitem{GMS} R.~Gopakumar, S.~Minwalla and A.~Strominger,
\textit{Noncommutative Solitons}, JHEP 0005 (2000) 020,
hep-th/0003160.




\bibitem{ABKM} I.Ya.~Aref'eva, D.M.~Belov, A.S.~Koshelev and P.B.~Medvedev,
\textit{Tachyon Condensation in the Cubic Superstring Field
Theory}, hep-th/0011117.

\bibitem{AMZ1} I.Ya.~Aref'eva,  P.B.~Medvedev and A.P.~Zubarev,
\textit{New representation for string field solves the consistency
problem for open superstring field}, Nucl.Phys.~B341 (1990) 464.

\bibitem{PTY} C.R.~Preitschopf, C.B.~Thorn and S.A.~Yost,
\textit{Superstring Field Theory}, Nucl.Phys.~B337 (1990) 363.



\bibitem{GrJe} D.~Gross and A.~Jevicki,
\textit{Operator Formulation of Interacting String Field Theory},
(I), (II), Nucl.Phys.~B283 (1987) 1;
Nucl.Phys~B287 (1987) 225.


\bibitem{GrJe3} D.~Gross and A.~Jevicki,
\textit{Operator Formulation of Interacting String Field Theory
(III). NSP superstring}, Nucl.Phys.~B293 (1987) 29.


\bibitem{l'Clare} A.~LeClair, M.E.~Peskin and  C.R.~Preitschopf,
\textit{String field theory on the conformal plane I},
Nucl.Phys.~B317 (1989) 411.

\bibitem{l'Clare2} A.~LeClair, M.E.~Peskin and  C.R.~Preitschopf,
\textit{String field theory on the conformal plane II},
Nucl.Phys.~B317 (1989) 464.

\bibitem{Marino} M.~Marino, R.~Schiappa,
\textit{Towards Vacuum Superstring Field Theory: The Supersliver}, hep-th/0112231.
\end{thebibliography}
\end{document}